\documentclass[journal]{IEEEtran}
\usepackage[a4paper, lmargin=1.5cm, rmargin=1.5cm, top=2.5cm,bottom=2.5cm]{geometry}
\usepackage{mathptmx}
\usepackage{hyperref}
\usepackage{multirow}
\usepackage{float}
\usepackage{enumitem}
\usepackage{amsmath}
\usepackage[english,brazil]{babel}
\usepackage[utf8]{inputenc}
\usepackage{color} 
\usepackage{colortbl}
\usepackage{graphicx} 
\usepackage{url}
\usepackage{mathtools}
\usepackage{arydshln}
\usepackage{amssymb}
\usepackage[normalem]{ulem}
\usepackage{amstext}
\usepackage{amsfonts}
\usepackage{float}
\usepackage{graphics}
\usepackage{epsfig}
\usepackage{cite} 
\usepackage[table,xcdraw]{xcolor}
\usepackage{enumitem}
\newcommand\myeq{\stackrel{\mathclap{\footnotesize\mbox{def}}}{=}}
\newcommand{\doi}{doi:}
\renewcommand*{\doi}[1]{\href{http://dx.doi.org/#1}{doi: #1}}
\newcommand{\arXiv}{arXiv:}
\renewcommand*{\arXiv}[1]{\href{https://arxiv.org/abs/#1}{arXiv: #1}}
\AtBeginDocument{%

}
\ifCLASSINFOpdf
\else
\fi
\begin{document}
\onecolumn
$~~~~~~~~~~~~~~~~~~~~~~~~~~~~~~~~~~~~~~~~~~~~~~~~~~~~~~~~~~~~~~~~~~~~~~~~~~~~~~~~~~~~~~~~~~~~~~~~~~~~~~~~~~~~~~~~~~~~~~~~$ 		 JEL: $\mathbf{L11;~L13;~L60,~L62}$\\

\vspace{1.5cm}
\LARGE
\noindent
Industrial Concentration of the Brazilian Automobile Market and Positioning in the World Market\\

\normalsize
\vspace{0.8cm}
Zionam E. L. Rolim$^{\dagger}$, Rafaël R. de Oliveira$^{\dagger}$, and Hélio M. de Oliveira$^{\ddagger}$\\

Federal University of Pernambuco (UFPE), Recife, Brazil.\\
\indent
$^{\dagger}$Department of Economic Sciences, CSSA, UFPE.\\
\indent
$^{\ddagger}$Department of Statistics, CCEN, UFPE.\\

\line(1,0){500}\\

$\mathbf{Abstract}$ -- This paper surveys the evolution of industrial concentration of the Brazilian automotive market as well as its positioning in the worldmarket. Data available by OICA (International Organization of Motor Vehicle Manufacturers) were used to better understand the characteristics of the Brazilian market on the world stage. A cluster analysis algorithm (by the k-means technique) ranks Brazil with a concentration profile in a group of countries like US and South Korea, in contrast to countries such as Germany, Canada and Japan, or even France and Italy. It is rather usual to characterize the market structure through industrial concentration indices: we revisit CR ratios (concentration ratios), HHI (Herfindahl-Hirschman index), B (Rosenbluth index), and CCI (Horvath comprehensive concentration index). Data of Anfavea-Brazil (Associação Nacional dos Fabricantes de Veículos Automotores) were used to estimate these indices in the period 2012-2018 for the national automobile industry. The values obtained indicate that by 1998 the automotive sector was behaving as an oligopoly-differentiated. However, the values of more recent periods (particularly CR4) strongly indicate that the sector is currently moderately concentrated and is changing for a quasi-devolved market. 
\\

$\mathbf{Keyterms}$ -- economic activity of the automotive industry; industrial concentration levels; concentration ratio (CR4); Herfindahl-Hirschman index (HHI); automotive market share.\\

\line(1,0){500}\\

\section{INTRODUCTION}
% The very first letter is a 2 line initial drop letter followed
% by the rest of the first word in caps.
% 
% form to use if the first word consists of a single letter:
% \IEEEPARstart{A}{demo} file is ....
% 
% form to use if you need the single drop letter followed by
% normal text (unknown if ever used by the IEEE):
% \IEEEPARstart{A}{}demo file is ....
% 
% Some journals put the first two words in caps:
% \IEEEPARstart{T}{his demo} file is ....
% 
% Here we have the typical use of a "T" for an initial drop letter
% and "HIS" in caps to complete the first word.
% ``starter file''
\small
\IEEEPARstart{T}
\normalsize
he automobile sector is considered to be one of the most important industrial activities in the modern world \cite{Womack-Jones-Roos}. It is intended to analyze the strategies of the Brazilian automobile industry and especially the market concentration, taking into account the different factors that influence such a process. Thus, we seek to understand the importance of the sector, as well as its possibilities for expansion. The automobile industry has been ordinarily regarded as a differentiated-concentrated oligopoly \cite{Pudo}. The automotive sector is characterized by a concentrated market structure, with a relatively small number of large corporations accounting for the largest share of production and total sales. In this structure, competition occurs through product differentiation, creating niche-markets in order to secure a market share. The automobile industry, including car, buses, trucks, tractors and the like, is heavily subject to fluctuations in demand \cite{Abreu}, \cite{Baumgarten}. Concentration by industry is a function of the number of companies operating and their respective market shares in total sales \cite{Djolov}. Market concentration is also used as a measure of competition \cite{Curry-George}, \cite{Rossi}. In a broad sense, the term industrial concentration is understood as a process that consists in increasing the control exerted by large companies over economic activity \cite{Barbosa}. The industry with higher profits, consequently, greater capacity for internal capital accumulation. This can give more incentives for new investments, increase their market power, lowering the participation of competing industries  \cite{Filho}, \cite{Hashmi}, \cite{Luttmer}. It is proposed in this work:
\vspace{0.2 cm}
\begin{itemize}
\item Study of the behavior of the Brazilian  motorcars market and its positioning in the world scenario. 
\vspace{0.1 cm}
\item Analysis of degree of concentration ($CR_4$), Herfindahl Hirschman Index ($HHI$), Horvath Extended Concentration Index ($CCI$), calculated based on the participation of the automakers in Brazil.
\vspace{0.1 cm}
\item Investigation of the automotive market concentration in historical comparison and in the strategies of the Brazilian automotive industry, seeking a better understanding of the competitiveness of the automotive sector.
\end{itemize}
\vspace{0.2 cm}

\noindent
In the industrial mass production systems, such as the automotive sector, the concentration of vehicle manufacturing in the hands of a small number of organizations has been consolidated. In the 1960s, the three major US automobile companies accounted for 94 percent of that country's production. In Germany, four companies made 91\% of the vehicles. In France, 100\% of the production came from four companies, and in Italy, just a single company produced 90\% of the vehicles on its own  (\cite{Toffler-Alvin}). In any case, a concentration similar to Pareto's 80/20 rule seems to be obeyed \cite{Cirillo}. The Pareto Principle or the law of the vital few \cite{Persky}, \cite{Rodd}:
\vspace{0.1 cm}
\begin{quote}
\underline{Pareto rule}: \textit{ ``for many events, approximately 80\% of the effects come from 20\% of the causes.''}
\end{quote}

\vspace{0.1 cm}
With the uncontrolled inflation in the early 1990s \cite{Feijo-et_alli}, vehicle prices increased and so did financing constraints and ban on new consortia, favoring poor sales performance until 1992 \cite{Comin}. Due to concerns about the future of domestic production and the ability of this sector to generate trade deficits, a slow opening process was chosen, without much investment. The second cycle of investments refers to the expansion of the domestic market, starting in the second half of the 2000s \cite{Botelho}, \cite{Pudo}. It was motivated by the resumption of the positive employment and income trajectory in the Brazil, expansion of internal demand and improvements in financing conditions, especially from the decline in interest rates and strengthening of corporate financing operations by the National Bank for Economic and Social Development (BNDES) \cite{Barros-Castro-Vaz-Hupsel}, \cite{Barros-Pedro}. In the period from 2006 to 2013, the automobile tax rate was reduced (excise tax exemption) in order to increase their demand. In this regard, see \cite{daSilva}, \cite{Wilbert_et_alli}. An important feature that differentiates the Brazilian market from other markets is investments in the production of the so-called popular cars. Most automakers focus their production on popular cars. In the first half of 1997, this modality reached 61\% (popular) and 69\% (small) of domestic car sales. Since 1997, companies have stepped up the introduction of more sophisticated versions of small motorcars. Flex-fuel autos now represent about 90\% of all light-duty registrations. Just to put it chronologically, the years in which the major automakers started business in Brazil are compiled: Ford 1919, GM 1925, VW 1953, Fiat 1976, Toyota 1988, Kia Motors 1992, Honda 1996, Daimler-Chrysler 1998, Renault 1998, PSA Peugeot-Citroën 2001, Chery 2009, JAC 2011, Hyundai (2012), Jeep (Fiat-Chrysler Automobile) 2015. The empirical part of the paper seeks to understand the concentration process of the automobile industry through historical analysis and a comparative method between Herfindahl-Hirschman Index ($HHI$) and $CR_4$ Index, which are the two types of  concentration indices most commonly used in studies to determine market competitiveness.  \cite{Costa-Rosa}, \cite{Naldi}, \cite{Silva-Cavalari-Onofre-Corso}.
\section{WORLD SCENARIO AND EXPORTS}

\begin{figure}[!htbp]
\centering
\includegraphics[scale=0.38]{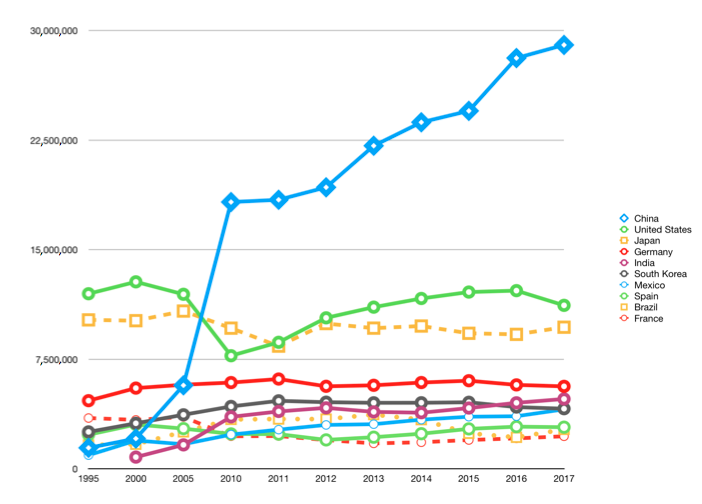}
\caption{{Evolution in world automobile production: the 10 largest producing countries (Brazil in 9th place). Source: OICA-\textit{International Organization of Motor Vehicle Manufacturers}.} }
\label{fig:evolucao}
\end{figure}

For a worldwide view, in Table \ref{tab:lista} we list the twelve largest automobile companies, and their productions (in thousand units per year) \cite{Sarti-Borghi}. The evolution in world automobile production, covering only the top ten producing countries, shows an evolution as identified in Figure ~\ref{fig:evolucao}. 
Data were obtained from the \textit{Organisation Internationale des Constructeurs d’Automobiles}, abbreviated by OICA (International Organization of Motor Vehicle Manufacturers). It was founded in 1919, and aggregates automobile industry associations from 39 countries \cite{OICA}. A focus on specific geographic areas of particular interest can be seen: in South America \cite{Santos-Pinhao}, or northeast Brazil  \cite{Gomes_Filho}. A noteworthy point is that China went from production of 2 million units in 2000 to more than 23 million units in 2014, increasing its share of world production from 3.5\% to 26.4\% in total in the period \cite{Barroso-Andrade}, \cite{Sarti-Borghi}.
A radical change in the global automobile market \cite{Barwick}.

\begin{table}[!hb]
\caption{Worldwide vehicle production by automaker (ranking of the 12 largest companies, 2014). 
Source: OICA.}
\label{tab:lista}
\begin{center}
\begin{tabular}{|c|c|}
\hline
automaker & total production (in thousand units)\\
\hline
Toyota & 10,475\\
Volkswagen & 9,895\\          General Motors & 9,609\\
Hyundai & 8,009\\
Ford & 5,970\\
Nissan & 5,098\\
Fiat & 4,866\\
Honda & 4,514\\
Suzuki & 3,017\\
PSA(Peugeot-Citroën) & 2,917\\
Renault & 2,762\\
BMW & 2,166\\
\hline
\end{tabular}
\end{center}
\end{table}

\begin{figure}[!htbp]
\centering
\includegraphics[scale=0.4]{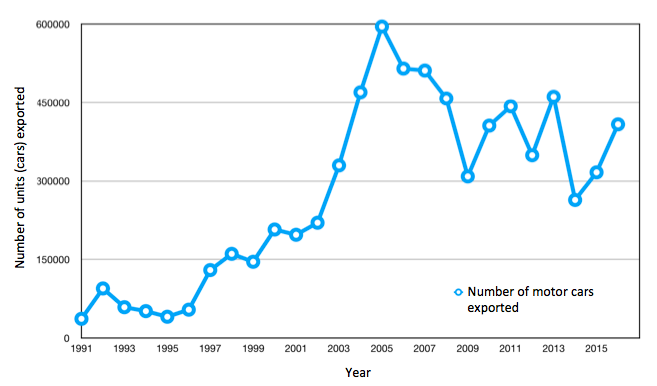}
\caption{Export volumes of  automobiles from Brazil, 1990-2017. Source: Anfavea. \url{http://www.anfavea.com.br/estat\%C3\%ADsticas.html}.}
\label{fig:exportacao}
\end{figure}

Table \ref{tab:concentra} explains the high degree of concentration, as ten companies account for 75\% of world production \cite{Rotta-Bueno}. In this evaluation the $CR_3$ and $HHI$  concentration indicators were calculated for several countries. Observing the $CR_3$ values in the different producing countries, a variation between [0.64, 1.00] is found. There is an  non-negligible correlation ($r^2\approx 0,65$) between these two indices in the case of the world motorcar industry. The crisis in internal vehicle sales has considerably increased domestic industry exports \cite{Carvalho}, \cite{Santos-Souza-Costa}. 
Brazil recorded a record export: 172,693 vehicles (ref. Q1 2017), a volume 69.7\% higher than in the same period for 2016 (Figure \ref{fig:exportacao}). For details on the main destinations of exports of automobiles made in Brazil, see Brazilian automobile industry directory 2017.
\begin{table}[!ht]
\caption{Indicators of car production concentration in the main producing countries in 1992. Source: \cite{Rotta-Bueno}. }
\begin{center}
\begin{tabular}{ccccc}
\hline
country     & \# automakers & total production & $CR_3$  & $HHI$  \\
\hline
Japan    & 9                    & 9,378,694      & 0.64 & 0.19 \\
USA      & 9                    & 7,082,000      & 0.90 & 0.33 \\
Germany & 7                    &     4,863,721      & 0.67 & 0.20\\ 
France & 2                    &     3,329,490      & 1.00 & 0.50\\ 
Spain & 9                    &     1,790,615      & 0.70 & 0.28\\ 
Italy & 8                    &     1,476,627      & 0.92 & 0.58\\ 
South Korea & 6                    &     
1,306,752      & 0.91 & 0.37\\ 
United Kingdom & 8               &  1,291,880      & n.d. & n.d\\ 
Canada & 6                    &  1,034,197      & 0.77 & 0.27\\
EU (Europe Union) & 6                       &  930,000      & n.d. & n.d. \\
Brazil & 4                       &  815,959      & 1.00 & 0.37 \\
Mexico & 4                       &  788,599      & n.d. & n.d. \\
\hline
\label{tab:concentra}
\end{tabular}
\end{center}
\end{table}

\section{CONCENTRATION INDICES}
\subsection{Concentration ratios CR}
The indices referred to as concentration ratios are established from the descending ordering of the studied variable \cite{Resende}. Thus, the share of the largest companies in total is considered, i.e. the ratio of the $k<n$ largest firms in a market with $n$ firms would be defined by \cite{Curry-George}:
\begin{equation*}
CR_k \myeq ~ \frac{\sum_{i=1}^k X_i}{\sum_{i=1}^n X_i},
\end{equation*}
where $X_i$ denotes the variable of interest to measure the degree of concentration. Alternatively, one can calculate
\begin{equation*}
CR_k \myeq ~ \sum_{i=1}^k P_i,
\end{equation*}
where $P_j \myeq ~ X_j/\sum_{i=1}^n X_i$ 
is the market share for the $jth$ firm.\\

Even though the value of $k$ is arbitrary, for the sake of simplicity, one chooses to work only with the participation of the four (respectively eight) largest companies. The respective indices are denoted by $CR_4$ e $CR_8$. These measurements are easy to compute because billing, installed capacity or sales information is usually available. The main shortcomings in the use of these indices are:
\begin{enumerate}
\item
In a given evaluation period, the $k$ largest companies considered may not be exactly the same as those considered for other periods.

\item
This approach disregards the relative concentration between firms, either within the group of the largest firms or even among the others.
\end{enumerate}

In Item 2, it is worth mentioning that the mergers that occur within the groups of  $n-k$ firms will not be captured, nor will the changes in the relative participation of each firm belonging to the group of largest $k$ be considered.

\subsection{Herfindahl-Hirschman Index (HHI)}
The index $H$ (or alternatively, $HHI$) was defined by the expression  \cite{Brezina}, \cite{Rhoades}
\begin{equation*}
H ~ \myeq ~ \sum_{i=1}^n P_i^2,
\end{equation*}
where $P_i ~ \myeq ~ X_i/\sum_{j=1}^n X_j$ accounts for the marketshare corresponding to $i$-th firm in total. This index was inspired by the mean square error rate introduced by Gauss and widely used in all sciences. Corresponds to the use of a quadratic weighting factor. In general, the expression could be put in the form
\begin{equation*}
H ~ \myeq ~ \sum_{i=1}^n W_i P_i,
\end{equation*}

\noindent
where $W_i$ is the weighting factor for each plot $P_i$. The particular choice where the weighting factor is $W_i=P_i$ leads to the square in terms, assigning greater weight the larger the plots. One could interpret the contributions of all $n$ firms described by a vector in Euclidean space, $\underline{P}~ \myeq ~ (P_1,P_2, \cdots, P_n)$ and since each coordinate is a contribution to the total, $\sum_{i=1}^n P_i =1$. Besides, $P_i \geq 0$.\\

\subsubsection*{Herfindahl Index Bounds:}

In the case of a monopoly, there is a firm  $P_1$ (without loss of generality) that is responsible for the entire market share: $P_1>0$ and for the other firms, $j \neq i$, $P_j=0$. The contribution is upper bounded by the unit:

\begin{equation*}
H = \sum_{i=1}^n P_i^2 \leq 1.
\end{equation*}

In order to examine the lower bound for $H$ (also denoted by $HHI$), we consider the Lagrangian

\begin{equation*}
\mathbb{L} \myeq ~ \sum_{i=1}^n P_i^2 - \lambda \left (  \sum_{i=1}^n P_i-1 \right ).
\end{equation*}
By imposing a null partial derivative of the Lagrangian, 

\begin{equation*}
\frac{\partial \mathbb{L}}{\partial P_j}=2P_j-\lambda=0,
\end{equation*}
it is founded $P_j$ as a function of the Lagrange multiplier, and by imposing the constraint $\sum_{i=1}^n P_i=1$,
one finds that $\forall j,~~P_j =\frac{1}{n}$. 
Straightforward to verify by the second derivative signal that the point is minimum. Combining the two bounds results for competition with $n$ firms in total gives the following: 

\begin{equation*}
\frac{1}{n} \leq H \leq 1.
\end{equation*}

The upper bound indicates the case where one company has maximum market power (monopoly). Looking at the lower bound, it can be seen that as the number of firms increases, the lower limit of the Herfindahl index will decrease. Obviously, when the number of firms tends to infinity, the index value tends to zero: $\lim_{n \to \infty} H=0$ (excluding monopoly).\\
Often, the $H$ index is considered expressed as points, where $P_i$ is a percentage: in this case, the bounds correspond to

\begin{equation*}
\frac{10,000}{n}  \leq H \leq 10,000 \text{ points}.
\end{equation*}

If there is a monopoly, the company has 100\% of the market and the upper bound is reached. According to the U.S. Department of Justice \& Federal Trade Commission \cite{Carlton}, \cite{Department_of_Justice}, the different scenarios are:\\
\begin{itemize}
\item \textit{Unconcentrated Markets}: $HHI$ below 1,500
\item \textit{Moderately Concentrated Markets}: $HHI$ between 1,500 and 2,500
\item \textit{Highly Concentrated Markets}: $HHI$ above 2,500
\end{itemize}
Some simple rules for the competition system can be established in terms of $HHI$. Here, the sub-indexes \flqq 0\frqq \ and \flqq 1\frqq \ means, before and after fusion, respectively. With $\Delta H \myeq ~ H_1-H_0$,

\begin{enumerate}[label=(\alph*)]
\item
If $H_1<0.1$ (market remains devolved.)
\item 
If $0.1 \leq H <0.18$ e $\Delta H<0.1$ (with fusion there was a small increase in concentration).
\item
If $H_1 \geq 0.18$ e $\Delta H<0.005$ (there is no causal nexus: the market was already concentrated).
\end{enumerate}
\subsection{Dominance Index (DI)} 
Again, it is assumed that there are $n$ companies operating in the market and the concentration function $F$, for some parameter $ \alpha>0$ exogenous to the model is
\begin{equation*}
F(Q,\alpha) ~ \myeq ~ \sum_{i=1}^n  \left ( 
\frac{P_i^{2\alpha}}{ \left ( \sum_{j=1}^n P_j ^{\alpha} \right )^2} 
\right ).
\end{equation*}
The cases of interest are \cite{Ten-Kate}
\begin{enumerate}
\item
for $\alpha=1$, $F(Q,1)=HHI$.
\item
for $\alpha >1$ there is an index of $\alpha$-dominance.
\end{enumerate}

The most commonly used case, $F(Q,2)=ID$, is called $DI$ and refers to \textit{dominance index}.

\begin{equation*}
DI ~ \myeq ~ \sum_{i=1}^n  \left ( 
\frac{P_i^4}{ \left ( \sum_{j=1}^n P_j ^2 \right )^2} 
\right ).
\end{equation*}

\subsection{Rosenbluth Index (B)}

The Rosenbluth index ($B$) is defined by taking into account that companies in an industry are ranked in such a way that $P_1 \geq  P_2 \geq P_3 \geq ... \geq P_k$.
through 
\begin{equation*}
B ~ \myeq ~ \frac{1}{2 \sum_{i=1}^k P_i-1}   
\end{equation*}
Again, when the industrial sector consists of a single company, $B=1$, i.e., the Rosenbluth index reaches its maximum value. On the other hand, this index approaches zero when production is relatively evenly divided by a large number of companies \cite{Meilak}.

\subsection{Horvath Comprehensive Concentration Index (CCI)}

Unlike $CR_k$ which is a measure of absolute concentration, the $HHI$ index is a relative measure of concentration. Measuring something midway between $CR_k$ and $HHI$ is the Horvath index (\cite{Horvath}), whose definition is:
\begin{equation*}
CCI ~ \myeq ~ P_1+\sum_{i=2}^n P_i^2 \times [1+(1-P_i)],
\end{equation*}
where $P_1$ is the market share of the largest holding firm (largest firm's share). \\

Given that the $CCI$ combines an absolute and relative concentration measure as compared to  $CR_k$ (which is an absolute measure) and $HHI$ (which is a relative measure), the values are expected to comply with inequality
\begin{equation*}
HHI \leq CCI \leq CR_k.
\end{equation*}

\section{BRAZILIAN AUTOMOBILE INDUSTRY IN THE WORLD}
Here, the idea is to visualize the behavior of Brazilian vehicle manufacturing groups in relation to the worldwide scenario \cite{Santos},  \cite{Scavarda-etalli}. An exploratory data analysis was performed by plotting the box diagrams (Tukey's box plot) for the concentration indices \cite{Campos}. 

\begin{figure}[!htbp]
\centering
\includegraphics[scale=0.4]{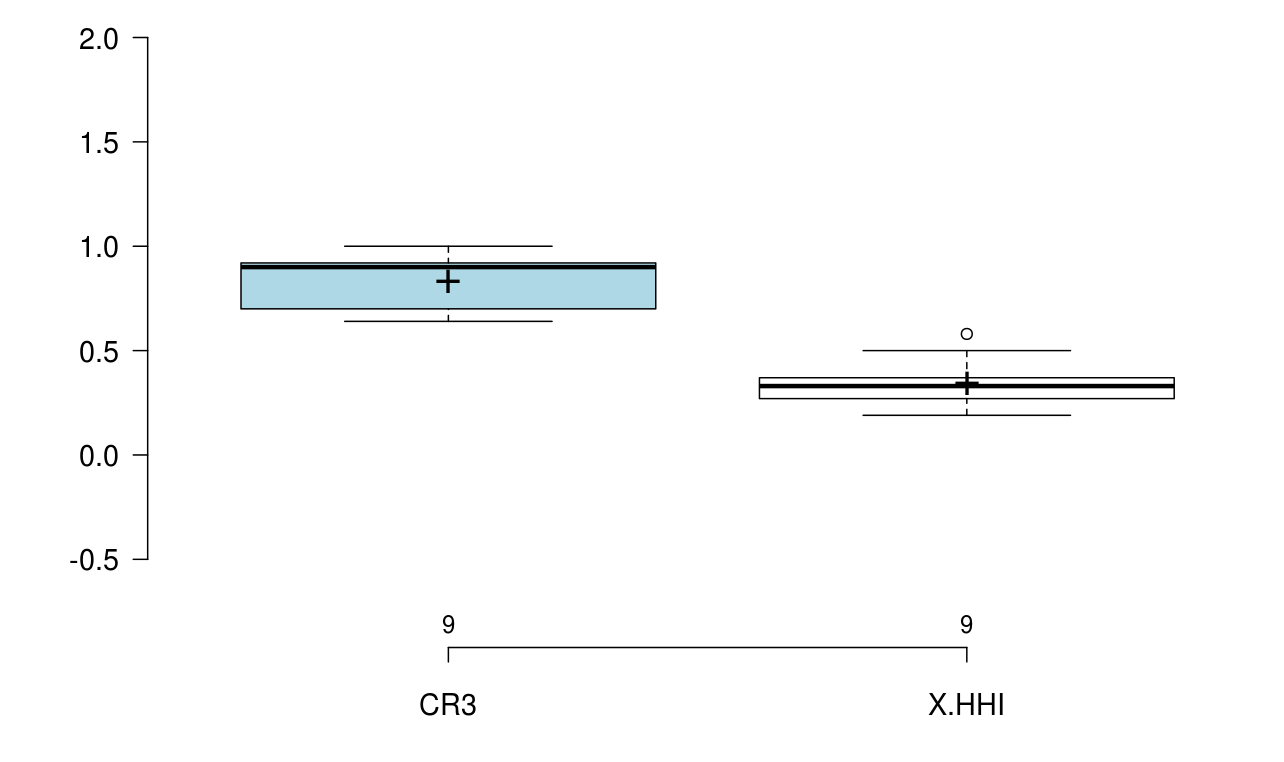}
\includegraphics[scale=0.4]{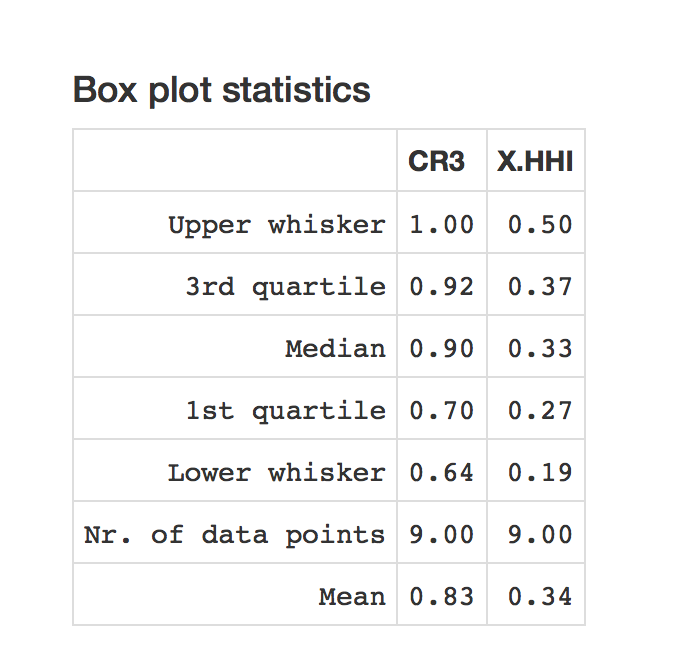}
\caption{{\textit{Boxplot} for $CR_3$ and $HHI$ concentration indices of automotive  industries in several countries.}}
\label{fig:boxplot}
\end{figure}

The graphs were obtained using the excellent free application available at \url{http://shiny.chemgrid.org/boxplotr/} and are shown in Figure \ref{fig:boxplot}. There is a positive asymmetry in the distribution of $CR_3$ indices, which does not occur for $HHI$. The boxplot statistical data are shown in Figure \ref{fig:boxplot}. Note that there is just a single outlier with respect to the $HHI$ index, which corresponds to \textit{Italy}. The explanation is obvious: Fiat is responsible for the excessive industrial concentration in that country, not comparable to what happens elsewhere ...

\begin{figure}[!ht]
\centering
\includegraphics[scale=0.47]{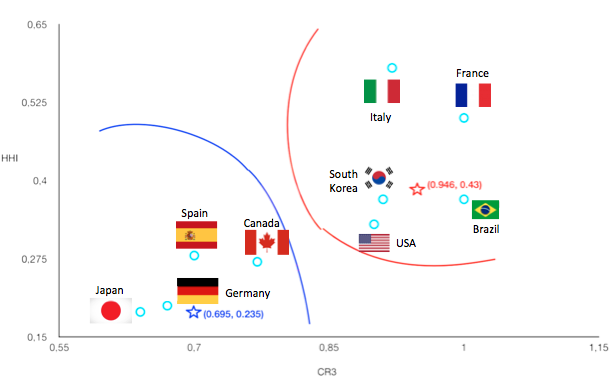}
\caption{{Cluster Analysis: Agglomerative classification between different countries according to the $CR_3$ and $HHI$ indices. Analysis in $k=2$ groupings.}}
\label{fig:2-Means}
\end{figure}

\begin{figure}[!htbp]
\centering
\includegraphics[scale=0.47]{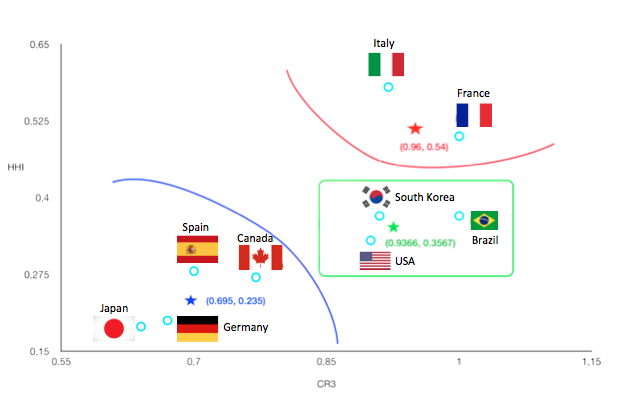}
\caption{{Cluster Analysis: Agglomerative classification between different countries according to the $CR_3$ and $HHI$ indices. Analysis in $k=3$ groupings.}}
\label{fig:3-Means}
\end{figure}

Another concern is the fact that both indices maintain some collinearity as they tend to estimate -- to some extent -- market concentration. For example, in the sector of interest (the automobile industry in various countries of the world, 2018), in the assessment of the $CR_3$ and $HHI$ concentration indices, a non-negligible correlation ($r^{2}=0.6489$) between them, fitted via linear regression, is obtained as 
$HHI=0.7434  CR_3-0.277$. 
The use of a cluster analysis algorithm, by the \textit{k-means} technique \cite{deOliveira}, results in the clusters displayed in Figures \ref{fig:2-Means} and \ref{fig:3-Means}. In the first, the country motor vehicle manufacturing are separated into the groups:

\vspace{0.2 cm}
\indent
\textcolor {blue} {$GR_1$=\{Germany, Canada, Spain and Japan\}} 
\vspace{0.1 cm}
\\
\indent
\textcolor {red} {$GR_2$=\{Brazil, South Korea, USA, France, Italy\}}
\vspace{0.2 cm}
\\
\indent
The calculated baricenters for both clusters are shown in Figure \ref{fig:2-Means} and correspond to the pairs $(0.695,~0.235)$ and $(0.946,~0.430)$, respectively. They were obtained using the online analysis available at \small{
\url{https://calculator.vhex.net/post/calculator-result/k-means-clustering}}. \\
\small

The separation between the two ``large'' groups ($GR_1$ and $GR_2$) suggests that in the countries of the $GR_2$ group, in which Brazil is inserted, they present higher industrial concentration. The values obtained by choosing  $k=3$ groups result in baricenters of the three groupings given by $(0.695,~0.235)$, $(0.937,~0.357)$, and $(0.960,~0.540)$. Of the indicated values, it is in Japan that the automobile market presents the greatest deconcentration. Considering a classification defined in three groups, the separation occurs in the partition of group $GR_2$, generating a group \textcolor{red}{\{Italy, France\}} in which the industrial concentration is more marked; in the first, by the power of Fiat, in the second, by Renault's marketshare, and the new group \textcolor{green}{\{South Korea, USA, Brazil\}}.

\subsection{Automobile Marketshare}

\begin{figure}[!htbp]
\centering
\includegraphics[scale=1.8]{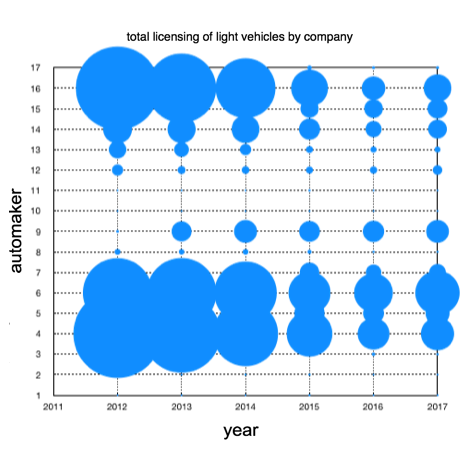}
\includegraphics[scale=0.5]{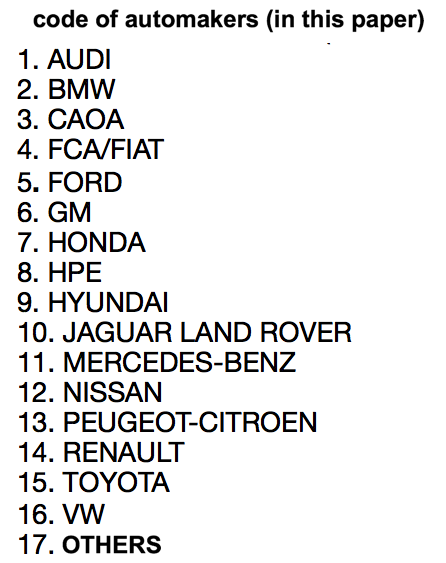}
\caption{Total licensing of light-duty vehicles by automaker: evolution between 2012 and 2017. Data Anfavea annuals \cite{ANFAVEA1}}
\label{fig:bolha}
\end{figure}

A table containing the number of annual passenger cars licenses (for this type of vehicle only) was compiled as a base. Only companies affiliated with ANFAFEA were considered \cite{ANFAVEA1}, which is representative of automobile production. Add to the fact that new licensed vehicles constitute a significant fraction of the produced motorcars. There are 16 companies, some like the FCA that includes $\{Chrysler, Dodge, Fiat, Jeep\}$, or HPE that involves $\{Mitsubishi, Suzuki\}$, or CAOA that includes $\{Hyundai, Subaru\}$ or Toyota, $\{Toyota, Lexus\}$.\\
\indent
A first sketch shown in Figure ~\ref{fig:bolha}, consists of a bubble chart. In it, the axles are: $X$ (year), $Y$ (\#automakers code), $size$ (Here, we chose to use the diameter of the bubbles proportional to the number of vehicles per \#automakers). 
Visual inspection (Fig. \ref{fig:bolha}) shows that industrial concentration has been decreasing in recent years. The behavior identified by the three large bubbles (labels 4, 6 and 16; FCA, GM and VW) is no longer so remarkable. The average values of light-duty vehicles licensed by automaker, standard deviations and coefficients of variation were calculated and are presented in Table \ref{tab:ESTATISTICA}.

Highlighted labels are shown in blue color. Car-makers that have been experiencing large fluctuations in production (with uneven annual production and more prone to crisis effects) include: Audi (+), FCA / FIAT (-), Hyundai (+), Peugeot-Citroën (-), and VW (-). Those of more stable production are Honda, Jaguar Land Rover and Nissan. Of the major automakers in Brazil, GM and Renault show slightly more regular production. Fiat and Volkswagen had a substantial reduction in production and sales, practically reduced to one third of 2012 levels. During this period, Hyundai went from a production of 20,000 units/year to almost 200,000 units/year. These data allow us to estimate, although not very accurate, the companies' market share. The justification is the high correlation between the number of annual licenses of passenger cars and the annual production of each automaker. Aiming to shorten the gap between the intention and effectiveness of a purchase, automakers are increasingly investing in marketing. Advertising investment (TV, Internet, newspapers and magazines) has been growing at a sustainable rate of over 10\% per year. Such a growth shows that the market is increasingly fierce among the major vehicle manufacturing groups. And wins who uses the best strategies, diversification and differentiation in each segment  \cite{Almeida-et_alli}, \cite{Rotta-Bueno}.
\footnotesize
\begin{table}[!htbp]
\caption{Exploratory analysis of  passenger vehicle licensing data (2012-2017) by automakers in the country. Data: Anfavea Yearbooks \cite{ANFAVEA1}.}
\label{tab:ESTATISTICA}
\begin{center}
\begin{tabular}{|lccc|}
%\begin{tabular}{p{16cm}} \hline
\hline
automaker & mean &standard deviation &CV\\
\hline
AUDI &10,533 &4,477	&43\%\\
BMW	&14,553 &3,012 &21\%\\
CAOA &46,606 &21,213 &21\%\\
\textcolor{blue}{FCA/FIAT}&\textcolor{blue}{449,155} &187,443 &42\%\\
FORD &246,389 & 59,650&24\%\\
\textcolor{blue}{GM} &\textcolor{blue}{428,665} & 112,833&26\%\\
HONDA &136,520& 10,210&7\%\\
HPE	& 33,447&13,420&40\% \\
HYUNDAI	&145,042 &60,877&42\%\\
JAGUAR LAND ROVER &8,922 &1,334 &15\%\\
MERCEDES-BENZ &	11,849& 3,708&31\%	\\
NISSAN & 66,293& 12,283&19\%\\
PEUGEOT-CITROEN	&82,659 & 38,616&47\%\\
RENAULT	&184,972 & 43,106&23\%\\
TOYOTA & 134,537& 30,246&22\%\\
\textcolor{blue}{VW} & \textcolor{blue}{394,317}& 191,198&48\%\\
OTHERS & 44,879&20,824 &46\%\\
\hline
\end{tabular}
\end{center}
\end{table}

\begin{table}[!htbp]
\caption{
Annual marketshare  corresponding to each automakers, based on the number of licenses: 2012-2017.}
\label{tab:marketshare}
\vspace{-0.5 cm}
\begin{tabular}{p{16cm}} \hline
\end{tabular}
\end{table}
\small
\begin{figure}[H]
\centering
\includegraphics[scale=0.5]{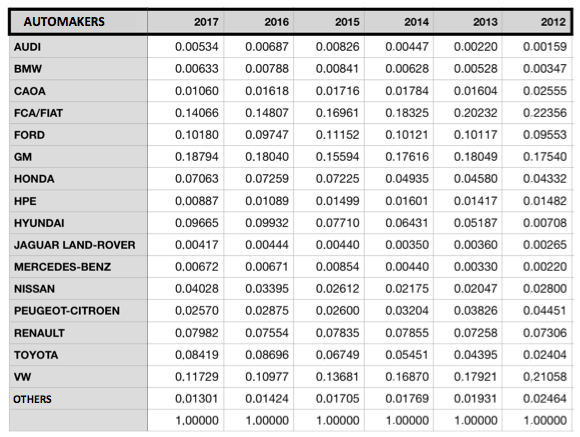}
\label{fig:marketshare}
\end{figure}

\subsection{Assessment of Concentration Indices (2012-2017) in the Brazilian Automotive Industry}

Following the formulas presented in the methodology section, and the data from Table \ref{tab:marketshare}, were calculated using a spreadsheet, the indices to quantify the industrial concentration in the automotive sector \cite{Pavic}. The results are summarized in Table \ref{tab:indices}. The fluctuation in the $CR_8$ index is less sensitive than that in $CR_4$, since it involves a larger number of automakers, conferring greater robustness.\\

\begin{table}[!htbp]
\caption{
$CR_4$, $CR_8$, $HHI$, $B$ and $CCI$ indices for the automobile industry in the 2012-2017 (table \ref{tab:concentra}).}
\label{tab:indices}
\begin{center}
\begin{tabular}{|l|ccccc|}
\hline
year	&$CR_4$	&$CR_8$	&$HHI$	&$B$ &$CCI$\\
	&		&		&		&		&\\
\hline
2012	&0.70508	&0.89396	&0.1463		&2.43809	&0.40003\\
2013	&0.66319	&0.87739	&0.1306		&3.06383	&0.36824\\
2014	&0.62931	&0.87603	&0.1215	&3.86668	&0.34670\\
2015	&0.57387	&0.86907	&0.1085	&6.76840	&0,31965\\
2016	&0.53755	&0.87011	&0.1072	&13.31406	&0.32160\\
2017	&0.54768	&0.87898	&0.1098	&10.48510	&0.32898\\
\hline
\end{tabular}
\end{center}
\end{table}

\begin{figure}[!htbp]
\centering
\includegraphics[scale=1.7]{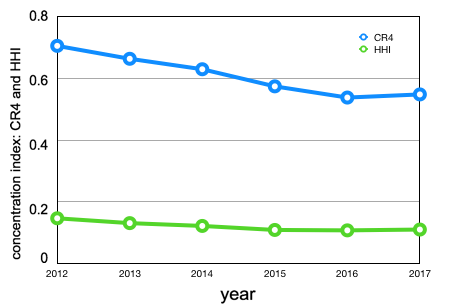}
\caption{
Evolution of concentration index estimates for $CR_4$  and $HHI$ for light-duty vehicles: 2012 to 2017.}
\label{fig:evolucao_indices}
\end{figure}

To facilitate redundant visualization, data on $CR_4$ and $HHI$  indices show a slow, but progressive deconcentration in the car industry. There is a slight increase in concentration in 2017, but in values that can be interpreted as statistical fluctuation and not a trend in the sector (Fig. \ref{fig:evolucao_indices}). The drop in the concentration index can probably be attributed to the installation of new automakers in the country, partitioning the motorcar market. Although automakers such as JAC (among others) were not included as they did not participate in ANFAVEA, part of the effect is reflected in the market reallocation. As a global assessment in the reference period (2012-2017), the following values of the median industrial concentration indices were calculated: $CR_4(median) \approx 0.6016$, $HHI(median) \approx  0.1156 $, $CCI(median) \approx 0.338$.

\small
\section{CONCLUDING REMARKS}
%\begin{minipage}[t]{\textwidth} 

It is possible to partially characterize the automobile market structure through indices of industrial concentration. Until 1998, there were only four companies operating in Brazil. With the entry -- around the year 2000 -- of five more automakers: Toyota, Renault, Mercedes Benz, DaimlerChrysler and Peugeot-Citroën \cite{Barros-Castro-Vaz-Hupsel}, it appears that the high concentration ratio in the sector has gradually declined as new companies were implemented in Brazil. Industrial concentration indices point out that the automotive sector -- which usually behaved as a differentiated-concentrated oligopoly, as it was characterized by a high degree of concentration (with over 90\% of the market being controlled by five manufacturers, namely: FCA/Fiat, General Motors, Volkswagem, Ford and Renault) -- has been changing \cite{Almeida-et_alli},  \cite{Ferraz-Almeida}, \cite{Silva_C}. The estimated concentration index values expressed via $CR_4$ up to 2010 were in a range that can be characterized as that of a highly concentrated market, clearly above 65\%. The recent estimates obtained in this paper do not cover all types of motor vehicles (excluding trucks, buses, agricultural machinery, wheel and grain harvester, crawler tractors, etc.). The focus was on annual all light-duty licensing, but there is of course a correlation with overall production in the country. Add to this the fact that only Anfavea-associated automakers were considered, excluding motor vehicle manufacturing groups such as JAC (China) or JEEP. This, of course, would naturally tend to reduce the concentration index estimates, both by increasing competitiveness and by the more distributed share of the 0 km light-duty motorcar market. In fact, it is no longer highly concentrated market. The values obtained for $CR_4$ (Table ~\ref{tab:concentra}) imply that the sector has moved from a highly concentrated market pattern to a moderately concentrated market ($45<CR_4<60$, \cite{Pavic}). This fact is striking in the period under review, contrasting with the characteristic profile observed until the beginning of the millennium. Prospects for the future seem to indicate a tightening of competition, an increase in the number of companies in the sector, with an expressive reduction in the sector's exaggerated industrial concentration. Regarding exports, from 2003 onwards, there was a significant increase in vehicle exports. Looking at the annual vehicle production variation data, the data available in Anfavea's statistical yearbook show that GM and Renault were the lowest-fluctuating installed companies in the period (2012-2017). In contrast, Fiat and VW showed (indirectly) the largest negative variations in production, the most directly affected by the crisis. Another interesting finding came from the analysis of Brazil's position worldwide in relation to the industrial concentration of the automobile industry. A cluster-based statistical study places Brazil with concentration profile in a group of countries such as the USA and South Korea, in contrast to the group of countries such as Germany, Canada and Japan, or even France and Italy. This view is interesting to boost the understanding of the behavior of the Brazilian automobile industry.
% use section* for acknowledgment
%\section*{Acknowledgment}
%The authors would like to thank...
% Can use something like this to put references on a page
% by themselves when using endfloat and the captionsoff option.
\ifCLASSOPTIONcaptionsoff
  \newpage
\fi
% references section
%
% <OR> manually copy in the resultant .bbl file
% set second argument of \begin to the number of references
% (used to reserve space for the reference number labels box)

\bibliographystyle{plain}
\renewcommand{\refname}{References}

% that's all folks
\end{document}